\documentstyle[12pt,epsfig,cite]{article}

\newcommand{\beq}{\begin{equation}}
\newcommand{\eeq}{\end{equation}}
\newcommand{\bea}{\begin{eqnarray}}
\newcommand{\eea}{\end{eqnarray}}

\newcommand{\gsim}{\lower.7ex\hbox{$
\;\stackrel{\textstyle>}{\sim}\;$}}
\newcommand{\lsim}{\lower.7ex\hbox{$
\;\stackrel{\textstyle<}{\sim}\;$}}

\def\lsim{\mathrel{\rlap{\lower3pt\hbox{\hskip0pt$\sim$}}
    \raise1pt\hbox{$<$}}}         
\def\gsim{\mathrel{\rlap{\lower4pt\hbox{\hskip1pt$\sim$}}
    \raise1pt\hbox{$>$}}}         

\renewcommand{\Im}{{\rm Im}\,}

\newcommand{\bibit}[1]{\bibitem{#1}}

\newcommand{\aver}[1]{\langle #1\rangle}

\newcommand{\Lam}{\Lambda_{\rm QCD}}

\newcommand{\GeV}{\,\mbox{GeV}}
\newcommand{\MeV}{\,\mbox{MeV}}
\newcommand{\matel}[3]{\langle #1|#2|#3\rangle}

\newcommand{\msp}[1]{\mbox{\hspace*{#1mm}~}}

\begin{document}
\thispagestyle{empty}
\vspace*{-10mm}

\begin{flushright}
Bicocca-FT-04-2\\
UND-HEP-04-BIG\hspace*{.08em}03\\
hep-ph/0403166
\end{flushright}
\vspace*{8mm}

\begin{center}
{\LARGE{\bf
Perturbative corrections to the semileptonic  \\
\boldmath $b\,$-decay moments:\vspace*{4mm} \\
{\Large{\bf $E^\ell_{\rm cut}$ dependence and running-$\alpha_s$
effects \vspace*{1.5mm}\\ in the
OPE approach}}
}}
\vspace*{10mm} 

\end{center}

\begin{center}
{\LARGE Nikolai~Uraltsev}\hspace*{1.5pt}\raisebox{5.5pt}{*} \vspace*{5mm} \\
{\sl INFN, Sezione di Milano,  Milano, Italy}\vspace*{.5mm}\\
{\small {\sf and}}\vspace*{.5mm} \\
{\sl Department of Physics, University of Notre Dame du Lac,
Notre Dame, IN 46556, USA}
\vspace*{10mm}

{\bf Abstract}\vspace*{-.9mm}\\
\end{center}

\noindent
We have calculated the perturbative corrections to all the structure
functions in the semileptonic decays of a heavy quark. Assuming an
arbitrary gluon mass as a technical tool allowed to obtain
in parallel all the BLM corrections. We report the basic applications,
viz.\ perturbative corrections to the hadronic mass and energy moments
with full dependence on the charge lepton energy cut. In the adopted
scheme with the OPE momentum scale separation around $1\GeV$ the
perturbative corrections to $\aver{M_X^2}$ are small and practically
independent of $E_{\rm cut}$; the BLM corrections are small, too. The
corrections to the second mass squared moment show some decrease with 
$E_{\rm cut}$ consistent with the effect of the Darwin operator,
within the previously estimated theoretical uncertainty. Perturbative
corrections in the pole-type schemes appear significant and vary with
$E_\ell$, decreasing the moments at higher cuts. The hardness of
hadronic moments is quantitatively illustrated for different cuts on $E_\ell$.

\setcounter{page}{0}

\vfill

~\hspace*{-12.5mm}\hrulefill \hspace*{-1.2mm} \\
\footnotesize{
\hspace*{-5mm}$^*$On leave of absence from 
St.\,Petersburg Nuclear Physics 
Institute, Gatchina, St.\,Petersburg  188300, Russia}
\normalsize

\newpage

\section{Introduction}

Inclusive decays of the heavy flavor hadrons are one of the most developed
dynamic uses of the fundamental QCD in the short-distance
regime. Application of the OPE allowed to address them at the
nonperturbative level through the expansion in the inverse mass of the
decaying quark: absence of the potentially largest 
${\cal O}(\frac{\Lam}{m_Q})$ corrections to the decay rates 
was established \cite{buv} and the
leading power corrections to inclusive decay distributions were obtained
through the expectation values of the local heavy quark operators in the
decaying hadron \cite{dpf,prl,koy}. As it has been known from the early
days of QCD \cite{banda}, the nonperturbative treatment is best
applied where usual perturbative corrections are comparable, or
subdominant to the leading nonperturbative effects. This environment
is often realized in the beauty decays once the OPE in the Wilsonian
implementation is applied, isolating effects of large distances even
from the traditional `perturbative' corrections. 

High precision checks of theory and practical use of the 
inclusive decay distributions are now performed with new data of better
reliability and of qualitatively different  statistics. Matching this
progress requires refinement of the theoretical
predictions. Perturbative corrections to a number of decay
distributions, first of all charge lepton spectrum in semileptonic
decays of  $b$ quarks, have been calculated long ago \cite{czarj},
along with many other characteristics \cite{gremst,fls95,flcut}. Most 
of the perturbative
calculations, although often rather sophisticated, however aimed at a
particular observable (total width, semileptonic spectrum or its
moments, angular asymmetries, etc.). As a result, in spite of the significant
intellectual effort invested, they often remained highly specialized,
without the possibility to be applied to only slightly modified 
observables.

The hadronic mass and energy distributions of the final state in the 
semileptonic decays appeared very promising for scrutinizing
nonperturbative QCD in the heavy quark system. In practice, experiments
typically have to apply lower cuts on the energy of the charged lepton, to
suppress backgrounds; this complicates the kinematics. Since, following
the first papers on the dynamic heavy quark expansion, the
power corrections were obtained directly for the semileptonic decay
structure functions \cite{koy,grekap}, calculating nonperturbative
effects for any inclusive moment became straightforward, more or less 
regardless of the lepton energy cut.

The perturbative effects, on the other hand, appeared in the role
of the weak link -- the perturbative corrections fully incorporating
the cut in the lepton energy have generally been unknown for hadronic
moments. 

To eliminate such an obstacle once and forever, we have calculated the
perturbative corrections directly to the semileptonic decay structure
functions. This allows straightforward evaluation of the perturbative
correction to all possible inclusive semileptonic
distributions. Moreover, we allowed for an arbitrary fictitious gluon
mass in the calculations; this opens the possibility to calculate
BLM corrections to an arbitrary order, or even to perform complete BLM
resummation similar to the one accomplished in
Refs.~\cite{blmvcb,imprec}, on the parallel footing. To be most 
universal for possible
applications, following Ref.~\cite{koy} we compute separately 
the structure functions induced by the vector and the axial-vector weak
current, as well as their interference ($w_3$). Likewise all five
structure functions are calculated, so the results can be readily 
applied to the semileptonic decays into $\tau$-leptons, etc. 
The explicit structure functions can also be used for computing
possible averages with the weight varying depending on the lepton
energy cut. This is one of the possibilities to optimize the trade-off
between the sensitivity to the heavy quark parameters and suppressing 
experimental backgrounds.

When this study was in the completing phase, the preprint by M.~Trott
\cite{trott} appeared, where the first results on the perturbative
structure functions proper were presented. This is clearly an important step 
in completing the 
theoretical toolbox required for precision analysis of $B$ decays. 
While addressing in general closely related subjects, we differ with
Ref.~\cite{trott} in a number of points calculation-wise, and also in
our applications. Aiming for ultimate flexibility in future
applications, we have calculated all five structure functions
separately for the vector and the axial-vector currents. Moreover, our
results are equally applicable to the first-order perturbative
corrections and to all BLM corrections.

Although having independent analytic calculations is
important for cross-checks, at this point we have not attempted to
compare our results with those reported in Ref.~\cite{trott}, even
for the pure one-loop limit. 
Programming alternative expressions anew for numerical evaluation
would represents a time-consuming process; it is not easily
safeguarded against possible typos at this technical, yet
unavoidable step.
To check our calculations we had used other opportunities mentioned
in Sect.~5, made available through our previous studies of
perturbative and power corrections in the semileptonic decays of heavy
flavors, or provided by the OPE applied to perturbation theory.

This paper reports a certain progress in the project presently carried
out together with Paolo Gambino, dedicated to the precision analysis
of the inclusive $B$ decays. It uses the Wilsonian implementation of the
OPE to combine perturbative and power-suppressed effects. In this
paper we mostly address the qualitative features and report only the
basic applications important for the ongoing experimental
analyses, which complement our recent publication \cite{slcm}.\footnote{Our
notations follow that paper; for the nonperturbative operators we
consistently use notations of Ref.~\cite{optical}.}
The comprehensive presentation is planned for the forthcoming 
paper \cite{future}.

\section{Inclusive \boldmath $B$ decays in the perturbative expansion}

The semileptonic structure functions are defined as the absorptive part of the
covariant structures appearing in the decomposition of the forward
scattering amplitude of the two weak currents off the $B$ meson:
\beq
h_{\mu\nu}(q^2\!, q_0) = \frac{1}{2M_B}
\matel{B}{\int {\rm d}^4 x \:{\rm e}^{-iqx}\:
iT\left\{J_\mu(x),J_\nu^\dagger(0)\right\} } {B}\;,
\label{10}
\eeq
where $J_\mu$ is generally the $\bar{c}\gamma_\mu b\,$ or/and 
$\,\bar{c}\gamma_\mu \gamma_5b$ current ($b\!\to \!u$ decay simply
corresponds to $m_c\!\to\!0$). Following the standard notations of
Ref.~\cite{koy} we put
\beq
h_{\mu \nu} = - h_1 g_{\mu \nu} + h_2 v_{\mu}v_{\nu} -
i h_3\epsilon _{\mu \nu \alpha \beta} v^{\alpha}q^{\beta}
+  h_4 q_{\mu}q_{\nu} +  h_5 (q_{\mu}v_{\nu} + v_{\mu}q_{\nu})\;,
\label{12}
\eeq
and 
\beq
w_i(q^2\!, q_0) = 2 \, \Im h_i(q^2\!, q_0) \;,
\label{q4}
\eeq
with $v_\mu$ denoting the 4-velocity of the decaying meson (or quark,
for perturbative calculations); $q^2$ and $q_0$ have the meaning of
the invariant mass squared and combined energy of the lepton pair, respectively.
All the decay distributions with light leptons are expressed, for
example in terms of the first three structure functions \cite{koy}:
\bea
\nonumber
\frac{{\rm d}^3 \Gamma}{{\rm d}E_{\ell\,}  {\rm d}q^{2\,} {\rm d}q_0 }
&\msp{-4}= \msp{-4}& \frac{G_F^2 |V_{cb}|^2}{32\pi^4}\,
\vartheta\!\left(q_0\!-\!E_\ell\!-\!\mbox{$\frac{q^2}{4E_\ell}$}\right)
\vartheta(E_\ell) 
\,\vartheta(q^2) \;\times \\ 
& & \msp{20} \left\{
2 q^2 w_1+[4E_\ell (q_0\!-\!E_\ell)\!-\!q^2]w_2 +
2q^2(2E_\ell\!-\!q_0) w_3
\right\} ;
\label{18}
\eea
the total integrated width without cuts on lepton energy depends only
on $w_1$ and $w_2$:
\beq
\Gamma_{\rm sl} \!=\! \frac{G_F^2 |V_{cb}|^2}{16\pi^4}
\int_{0}^{m_b^2} \!{\rm d}q^2   \int_{q_0>\sqrt{q^2}}^{m_b} {\rm d}q_0
\sqrt{q_0^2\!-\!q^2} \left(q^2 w_1(q^2\!,q_0) 
+\frac{1}{3}(q_0^2\!-\!q^2) w_2(q^2\!,q_0) \right)\;.
\label{20}
\eeq

Perturbative structure functions to order $\alpha_s^1$ consist of
virtual corrections $\propto
\delta(q_0-\frac{m_b^2+q^2-m_c^2}{2m_b})$, and bremsstrahlung
contributions with
$q_0<\frac{m_b^2+q^2-(m_c+\lambda)^2}{2m_b}$, where
$\lambda$ is the gluon mass. In the limit $\lambda^2 \!\to\!
0$ virtual corrections logarithmically diverge. The divergent part is
given precisely by the tree-level (free quark) structure functions, see
Ref.~\cite{koy}, Eqs.~(A1--A11), with the universal coefficient depending on
$q^2$. The real emission part at small $\lambda^2$ has a singularity
near the free quark kinematics, 
\beq
w_i(q^2, q_0) \propto \frac{1}{q_0-\frac{m_b^2+q^2-m_c^2}{2m_b}} \,,
\label{22}
\eeq
with the commensurate coefficient, so that the divergence cancels 
once integration over $q_0$ is
performed. At $q^2$ approaching $(m_b\!-\!m_c)^2$ the coefficient
vanishes, therefore shrinking domain of integration over $q_0$ does
not affect the cancellation of divergences. 

Perturbative structure functions themselves are strongly
infrared-sensitive in the dominant domain close to the free quark decay
kinematics. However, the integrals entering the experimentally
measured inclusive moments are not. Moreover, their infrared sensitivity is
governed by the OPE, and it is greatly reduced when using the
Wilsonian separation between perturbative and power corrections based
on the momentum scale a particular contribution originates
from. This separation is in practice done directly for the moments 
rather than for the structure functions themselves.

The analytic expressions for the ${\cal O}(\alpha_s)$ bremsstrahlung 
structure functions $w_k(q^2\!,q_0)$ are relatively simple containing at
worst $\log$s of the kinematics-related square roots, even at arbitrary
gluon mass. Yet they are lengthy for separate structure functions,
consisting of many terms at $\lambda^2\!\ne\! 0$. They simplify
significantly at $\lambda^2\!=\!0$. Virtual corrections at
$\lambda^2\!\to\! 0$ are well known one loop renormalization of quark
currents containing dilogs at worst. At arbitrary $\lambda^2$, however
the one-loop vertices with three different internal masses and a general
momentum transfer are too special functions. In practice, we represent
them as one-dimensional integrals over a single Feynman parameter $u$,
with $0\!\le\! u \!\le\! 1$.

Obtaining a general moment with the lower cut on lepton energy we, therefore
integrate the explicit structure functions over $q_0$ and $q^2$ (for
bremsstrahlung), or over $u$ and $q^2$ (for virtual corrections), with
the weight which generally is a polynomial in $(m_b \!-\!q_0)$ and
$(m_b^2+q^2-2m_b q_0-m_c^2)$. The lepton cut enters through the
concrete weight following from Eq.~(\ref{18}): it is obtained by
integrating the coefficients from $E^\ell_1$ to  $E^\ell_+$, where
\beq
E^\ell_{\pm}= \frac{q_0 \pm \sqrt{q_0^2\!-\!q^2}}{2}\;, \qquad 
E^\ell_1= \mbox{max}\{E_{\rm cut},\, E_-\}\;;
\label{24}
\eeq
the weights are simple polynomials in  $E^\ell_1$ or $E^\ell_+$.

\section{Applications}

Recent experiments at $B$ factories provide data of impressive
precision and quality, which allows for stringent tests of the heavy
quark expansion for inclusive decays, direct experimental extraction
of many heavy quark parameters and for robust defendable extraction of
$V_{cb}$ and $V_{ub}$ from the integrated decays rates. The current
precision requires to fully implement the applied cut on the charge lepton
energy in the theoretical calculations. In the recent publication
\cite{slcm} the
OPE-based predictions were given for various moments with cuts, in the
framework which
does not rely on assuming charm to be a heavy quark, but only expanding in
$1/m_b$. Perturbation theory-wise we used the Wilsonian approach which
assumes excluding soft gluon contributions from the coefficient functions. This
rendered the perturbative corrections small in size and presumably
stable against higher-order corrections.

For hadronic mass moments Ref.~\cite{slcm} evaluated 
the perturbative corrections
without the cut on lepton energy, since the full perturbative corrections
with cuts had not been known. Although some simpler parts of the
corrections had been calculated, including them would not be consistent:
in hadronic moments the terms proportional to different powers of
$M_B\!-\!m_b$ essentially mix under renormalization. In particular,
for the average hadronic mass square,
\beq
\aver{M_X^2}= m_c^2 + (M_B\!-\!m_b)^2 +  
2(M_B\!-\!m_b)\aver{E_x} + \aver{m_x^2\!-\!m_c^2} 
\label{28}
\eeq
the last term having the
meaning of the excess of the combined parton invariant mass at the
quark level over $m_c^2$, is
fed under renormalization by the preceding term driven by $\aver{E_x}$,
the average hadron energy at the parton level \cite{slcm}. 

The justification for approximation neglecting $E^\ell$-cut 
in the perturbative
corrections was, again, using the Wilsonian version of the
OPE. The absolute size of the perturbative corrections is
suppressed here, so even their noticeable variation with the cut was
not expected to produce a significant bias. Moreover, it is the soft
parton processes that are most sensitive to the particular
kinematics. The truly hard gluon effects contributing the perturbative
corrections in our approach, are expected to be less dependent on the
details of the kinematics as long as the process in question remains
sufficiently `hard'. 

\thispagestyle{plain}
\begin{figure}[hhh]\vspace*{-3.4mm}
\begin{center}
\mbox{\epsfig{file=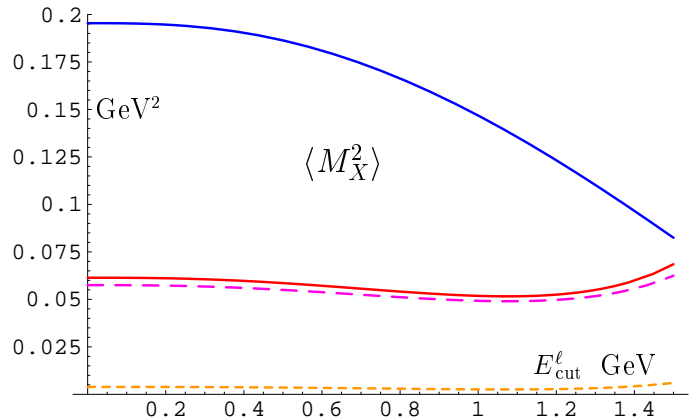,width=90mm
}}
\end{center}
\vspace*{-6.0mm}
\caption{ \small
Dependence of the perturbative corrections to $\aver{M_X^2}$ on  $E_{\rm
cut}$, for $\mu\!=\!1\GeV$ (red curve). Long-dashed and short-dashed curves show
the separate contributions of $\Delta_c$ and $E_x$, respectively. Blue
line (top) gives the perturbative correction in the pole-type scheme.
}
\end{figure}

With full perturbative expressions available, we can check validity
of those temporary assumptions. We indeed found that in our scheme
the perturbative corrections to $\aver{M_X^2}$ are practically
independent of $E_{\rm cut}$ in the whole domain $E_{\rm cut}\lsim 1.4
\GeV$. A more noticeable -- yet still not very significant --
variation with $E_{\rm cut}$ is observed for the second hadronic
moment with respect to average, $\aver{(M_X^2\!-\!\aver{M_X^2})^2}$. This
is expected, since this higher moment is more sensitive to the Darwin
operator, for which we did not actually remove the corresponding soft 
contributions from the leading-order coefficient function. The observed
 $E_{\rm cut}$-dependence, in particular the sign and the size are
compatible with such a contribution from the Darwin expectation value:
its coefficient function is negative and increases in magnitude
for larger $E_{\rm cut}$ \cite{slcm}, see, e.g.\ Table~6 of Ref.~\cite{slcm}.

The full  corrections are illustrated in Figs.~1 and 2 as functions of
$E_{\rm cut}$. For comparison we also show the corresponding effect in
the `usual' (pole-type) perturbative corrections without Wilsonian
separation. It is evident that for the first moment the effect is
dramatically different. 

\thispagestyle{plain}
\begin{figure}[hhh]\vspace*{-3.4mm}
\begin{center}
\mbox{\epsfig{file=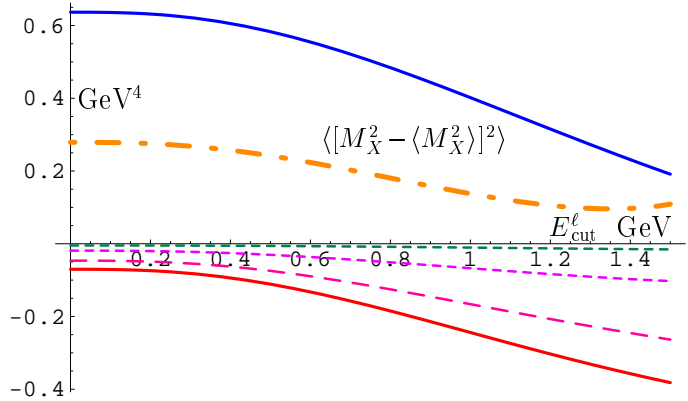,width=90mm
}}
\end{center}
\vspace*{-6.0mm}
\caption{ \small
Perturbative corrections to
$\aver{[M_X^2\!-\!\aver{M_X^2}]^2}$, 
for $\mu\!=\!1\GeV$ (red curve), and their breakdown showing 
the separate contributions  $\propto (M_B\!-\!m_b)^k$, 
$k\!=\!0,\,1,\,2$ (dashed curves). 
Blue line (top) refers to the pole
scheme. Orange dashed-dotted line shows subtracting the soft piece  of
$0.062\GeV^3$ of the Darwin expectation value; it would correspond to Wilsonian
$\rho_D^3$ normalized at $0.9\GeV$.
}
\end{figure}

\thispagestyle{plain}
\begin{figure}[hhh]\vspace*{-3.4mm}
\begin{center}
\mbox{\epsfig{file=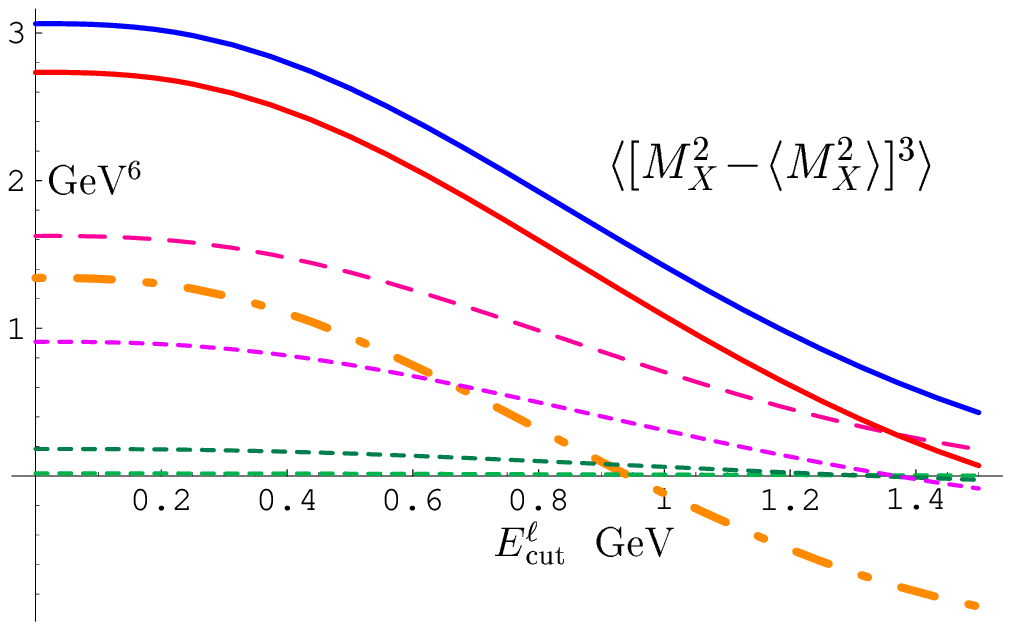,width=90mm
}}
\end{center}
\vspace*{-6.0mm}
\caption{ \small
Similar plot for the third invariant mass moment 
$\aver{[M_X^2\!-\!\aver{M_X^2}]^3}$. Total perturbative corrections are
reduced (orange dashed-dotted line) using $\rho_D^3(0.9\GeV)$ 
compared to the case of $\tilde\rho_D^3$ (red solid line). Blue solid
line corresponds to the pole scheme.
}
\end{figure}

The third moment of the hadronic invariant
mass squared  places, for the perturbative corrections, most weight on harder
gluons. In its soft part it
is mostly sensitive to the $D\!=\!6$ Darwin operator. Therefore, using the
Wilsonian prescription only for the effects scaling like $1/m_b$ and
$1/m_b^2$ does not make a noticeable difference, Fig.~3. The formal
perturbative contribution is then significantly dependent 
on the cut;\footnote{Ref.~\cite{slcm} evaluating the third hadronic
mass moment did not include the perturbative corrections to the terms
$\propto \!(M_B\!-\!m_b)^k$ with $k\ge 1$, even at zero cut. We have
computed them, and  numerically they turned out to be about $1\GeV^6$
(the dashed lines excluding the highest one),
decreasing fast from $E_{\rm cut}\gsim 500\MeV$.}
one should keep in mind, though that 
a significant fraction of it still comes from gluon momenta below
$1\GeV$. (For illustration we show the result of subtracting the 
contribution of the Darwin
expectation value of $0.062\GeV^3$ corresponding to $\mu\!=\!0.9\GeV$
in fixed-order perturbation theory). In practice,  the third moment
so far is important mainly as an estimate of the scale of $\tilde
\rho_D^3$; as had been pointed out \cite{amst,fpcp03} other unaccounted
effects introduce theoretical uncertainties of the same
magnitude.

\section{Running \boldmath $\alpha_s$ and BLM corrections}

We are also in the position to calculate 
the so-called BLM corrections -- the effects
accounting for running of the strong coupling $\alpha_s$ in one-loop
perturbative diagrams. Such effects are typically quite significant 
where the perturbative calculations do not eliminate explicitly the
low-momentum domain; however, they are moderate, or even small in size
in the appropriate Wilsonian scheme. A dedicated discussion for the
case of the total semileptonic width was given in
Refs.~\cite{blmvcb,imprec}. We leave a detailed implementation of this
method for the future publication \cite{future}, and here place the emphasis
on the qualitative features. They allow to assess the possible
accuracy of the theoretical predictions without much numerology.

\thispagestyle{plain}
\begin{figure}[hhh]\vspace*{-3.4mm}
\begin{center}
\mbox{\epsfig{file=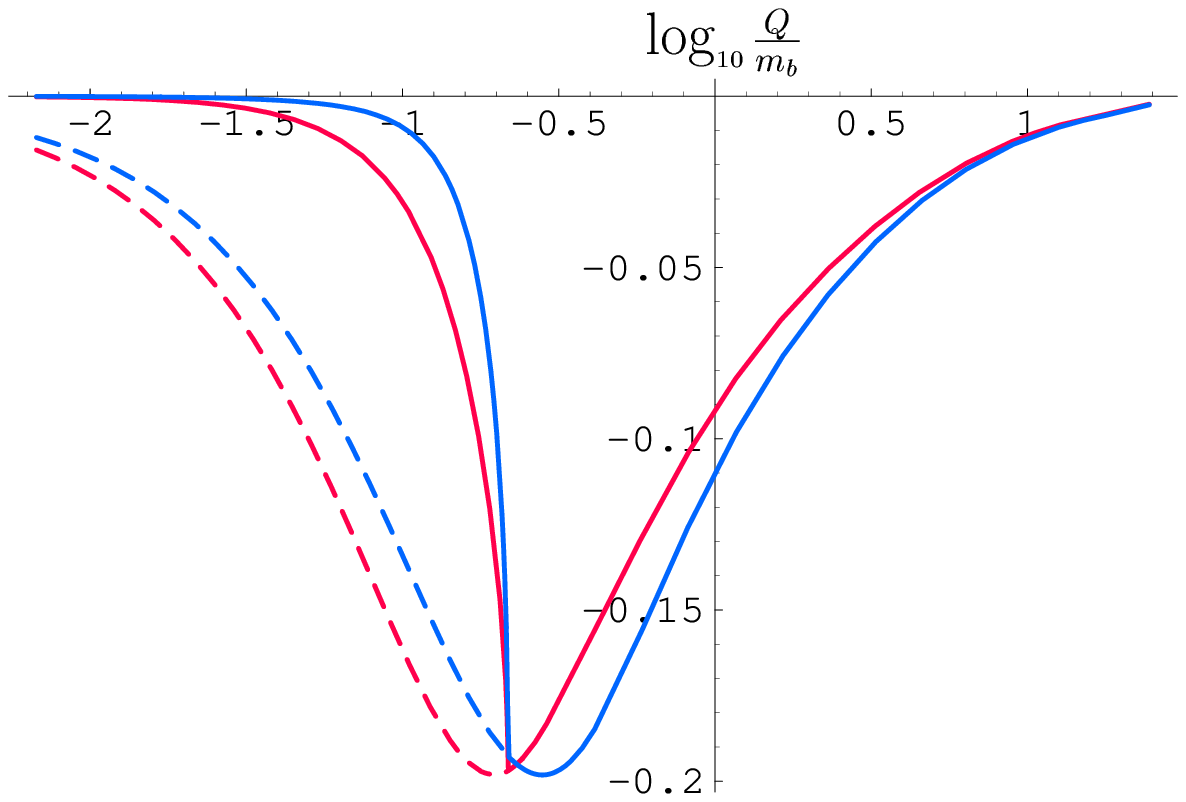,width=224pt
}}
\end{center}
\vspace*{-6.0mm}
\caption{ \small
Distribution over the gluon momentum for the total width without a cut
(blue) and for $E_{\rm cut}\!=\!1.5\GeV$ (red). Solid lines correspond to
the cutoff scale $\mu\!=\!1\GeV$, dashed lines show the unsuppressed 
contributions in the pole-type schemes. The area bounded by a curve
yields the overall first-order perturbative coefficient.
}
\end{figure}

The technique for the BLM summation in the Wilsonian approach was first
discussed in Ref.~\cite{blmope}; it is concisely presented in recent
Refs.~\cite{blmvcb,imprec}, and we do not describe it here. Neither we
detail numerical predictions or address all experimentally
interesting observables. We also largely leave aside more subtle 
theoretical aspects related to
full BLM resummation for higher moments with only a few lowest
nonperturbative operators included. 
Our main interest lies in the most sensitive case for theory,
viz.\ the width and the first hadronic moment $\aver{M_X^2}$ with the
lepton energy cut.

We start with the decay width itself when the lepton energy cut is
imposed. A convenient way to visualize the contribution of various
momentum scales $Q^2$ in one-loop perturbation theory for an observable $A$ 
is provide by the distribution
\beq
\frac{{\rm d} A^{\rm pert}}{{\rm d}\ln{Q^2}} = 
-\frac{{\rm d} A^{(1)}(\lambda^2)}{{\rm
d}\ln{\lambda^2}}\,\rule[-14pt]{.5pt}{30pt}\,
\raisebox{-10pt}{$_{\lambda^2=Q^2}$}\;,
\label{32}
\eeq
where $A^{(1)}(\lambda^2)$ is the first-order perturbative
coefficient calculated with the fictitious non-zero gluon mass
\cite{bbb6,dmw}. This representation has some limitations, yet is
quite suitable for qualitative purposes and in adopted in the
plots presented below.

Fig.~4 shows the distribution over the gluon virtualities for
the two extreme cases, $E_{\rm cut}\!=\!0$ (no cut) and $E_{\rm
cut}\!=\!1.5 \GeV$. Without the scale separation, dashed lines, the
perturbative corrections are clearly not too well behaved, since the
major contribution comes from the gluon momenta noticeably below
$1\GeV$. This is an artefact brought in by the pole masses
\cite{pole,bbz} used by conventional perturbative diagrams. Fig.~4 shows
that applying Wilsonian procedure quite effectively eliminates this
domain (solid line).

At first glance, there is no much difference between the perturbative
corrections for $E_{\rm cut}\!=\!1.5 \GeV$ compared to the total 
width without kinematic restrictions.  However, the low-momentum tail is
much higher for $\Gamma_{\rm sl}(E^\ell\!>\!1.5\GeV)$. This leads to a far
more significant impact of the soft physics, the fact deduced
independently from the growth of the Wilson coefficient for the
higher-dimension nonperturbative operators (e.g., Darwin). As
anticipated, the effect is more significant for moments of
the distributions. We will now look closer at the average hadron invariant
mass squared $\aver{M_X^2}$.

\thispagestyle{plain}
\vspace*{.3mm}
\begin{figure}[hhh]
\mbox{\epsfig{file=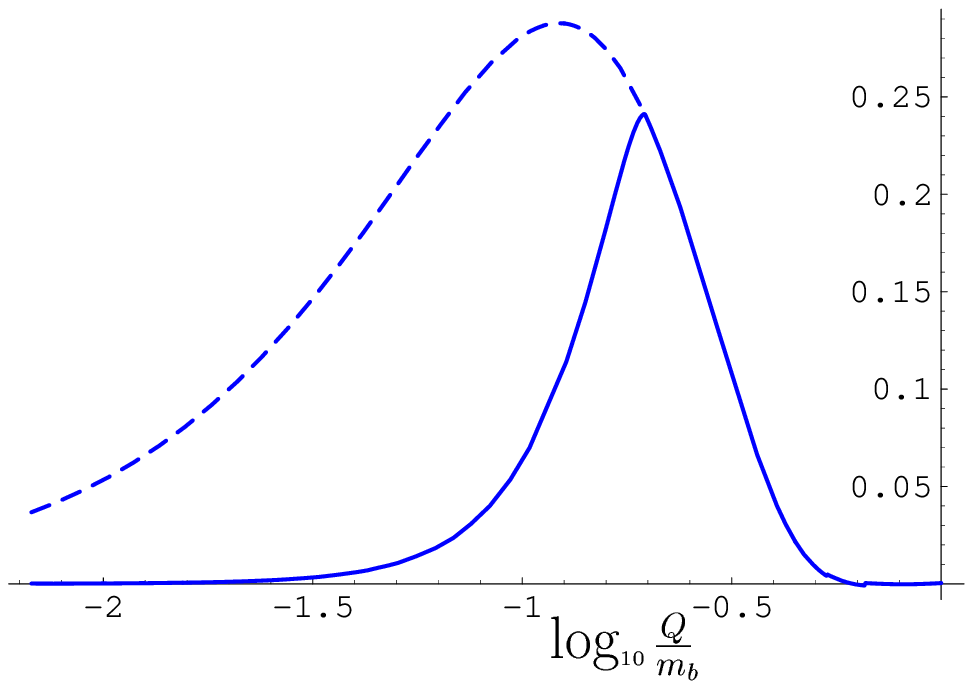,width=7.4cm}}
\hfill
\mbox{\epsfig{file=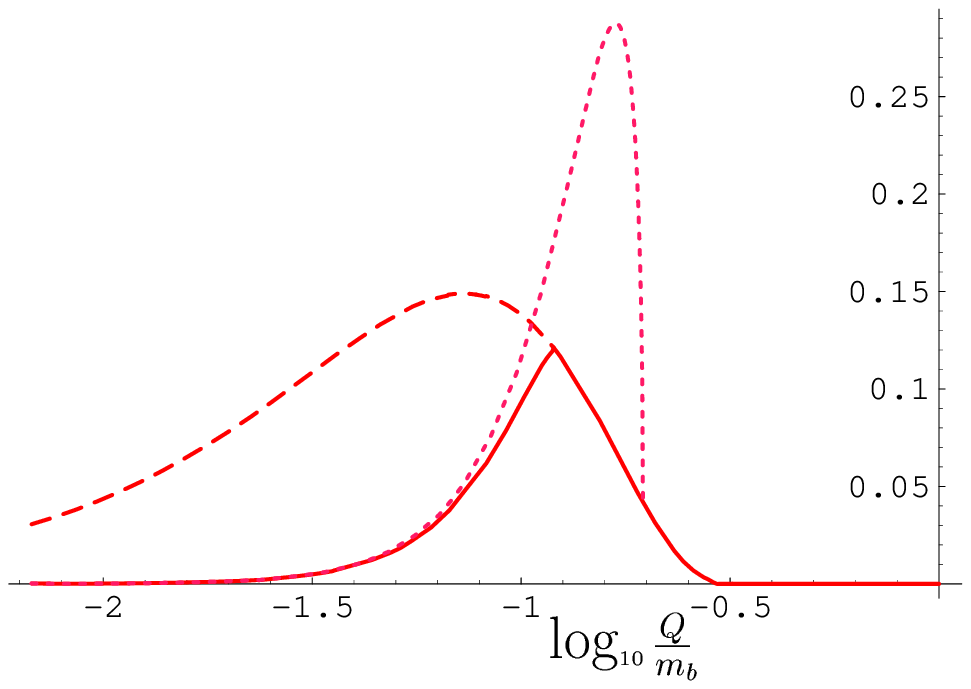,width=7.4cm}}\vspace*{-4.5mm}\\
\begin{minipage}[t]{7.4cm}
\caption{ \small
Similar distribution for the dominant perturbative contribution
$\Delta_c$ to
$\aver{M_X^2}$, without lepton energy cut. Solid line is for
$\mu\!=\!0.9\GeV$, dashed for $\mu\!=\!0$ (pole scheme).
}
\end{minipage} \hfill
\begin{minipage}[t]{7.4cm}
\caption{ \small
The case of $E_{\rm cut}\!=\!1.5\GeV$. Dotted line refers to the same
separation scale $\mu\!=\!0.9\GeV$; solid line shows the 
choice of $\mu\!=\!0.55\GeV$.
}
\end{minipage} 
\end{figure}

In the OPE, $M_X^2$ consists of two distinct dynamic pieces (they 
mix under the renormalization of the effective low-scale QCD
for heavy quark):
\beq
\aver{M_X^2}=m_c^2+(M_B\!-\!m_b)^2+ (M_B\!-\!m_b)\aver{2E_x} +
(m_x^2\!-\!m_c^2)
\label{40}
\eeq
\vspace*{-8mm}
$$
E_x\equiv m_b\!-\!q_0\;, \qquad \Delta_c=m_x^2\!-\!m_c^2 
\equiv m_b^2+q^2\!-\!2m_b q_0\!-\!m_c^2\;.
$$
The first one, $(M_B\!-\!m_b)\aver{2E_x}$ dominates corrections
to $\aver{M_X^2}$ \cite{WA}. However, among the perturbative corrections those to
$\aver{\Delta_c}$ appear most significant. Therefore, we start with
the second piece, $\aver{\Delta_c}$.

At zero lepton cut, $\Delta_c$ which in the conventional perturbative
diagrams is described by real gluon emission, is very soft,
see Fig.~5: the four-body phase space for $b\!\to\!cg+\ell\nu$ is very
sensitive to the gluon mass even when the latter is only a few hundred
$\MeV$. Presence of such corrections would be a
disaster for precision calculations. However, in QCD the phase 
space-unsuppressed {\sf soft}
gluon emissions feeding $m_x^2$ are related by gauge symmetry to the
Coulomb self-energy of the heavy quark \cite{optical}. The Wilsonian
treatment then effectively eliminates the infrared domain, as
illustrated by the solid curve corresponding to the separation
scale $\mu\!=\!0.9\GeV$.

The similar distribution for the cut at $1.5\GeV$ is shown in
Fig.~6. Applying the same Wilsonian cutoff at $0.9\GeV$ we eliminate
the deep infrared domain as well, yet paying the price of quite
unphysical spike in the contribution of gluons at the scale around
$1\GeV$. It disappears only when $\mu$ is lowered down to $0.55\GeV$. 
This behavior does not mean that the masses and
higher-dimension operators normalized at, say $1\GeV$ cannot be used
for perturbative calculations. The perturbative result can be
expressed in terms of any short-distance masses, for instance $\bar{m}_b(m_b)$
and $\bar{m}_c(m_b)$ corresponding to $\mu$ in a few $\GeV$
range. It rather demonstrates that the effect of the domain of momenta
between $600\MeV$ and $1\GeV$ is not fairly described by only the
leading $1/m_b$ and $1/m_b^2$ contributions -- the terms scaling like
higher powers of $1/m_b$ are comparable, or even dominate them. This
is a direct consequence of the deteriorating hardness of the inclusive
width with the high cut, which effectively introduces another, much
lower than $m_b$ mass scale parameter in the OPE \cite{amst,fpcp03,
misuse}.

\thispagestyle{plain}
\begin{figure}[hhh]\vspace*{-3.4mm}
\begin{center}
\mbox{\epsfig{file=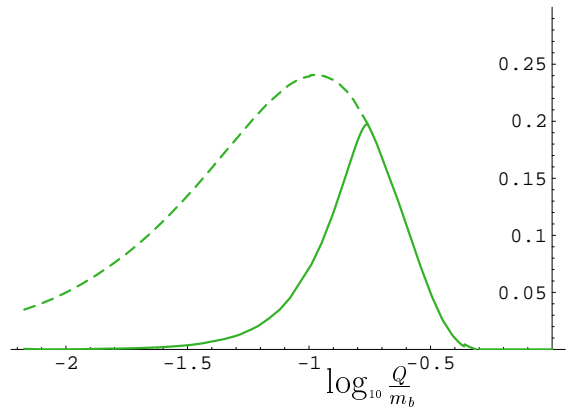,width=74mm
}}
\end{center}
\vspace*{-6.0mm}
\caption{ \small
The case of a mild cut at $E^\ell\!=\!0.9\GeV$, using $\mu\!=\!0.8\GeV$.
}
\end{figure}

For milder cuts with lower cutoff energy the convergence of the power
expansion improves; for instance, Fig.~7 shows the similar
distribution at $E_{\rm cut}\!=\!0.9\GeV$ where it still
looks perfect at $\mu\!=\!0.8\GeV$.

\thispagestyle{plain}
\vspace*{.3mm}
\begin{figure}[hhh]
\mbox{\epsfig{file=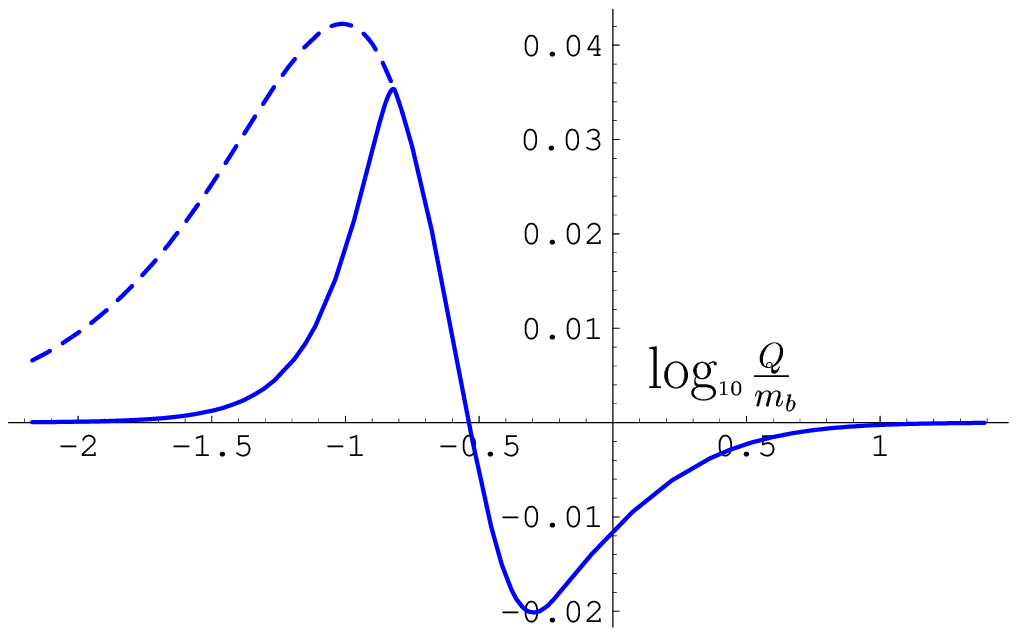,width=7.4cm}}
\hfill
\mbox{\epsfig{file=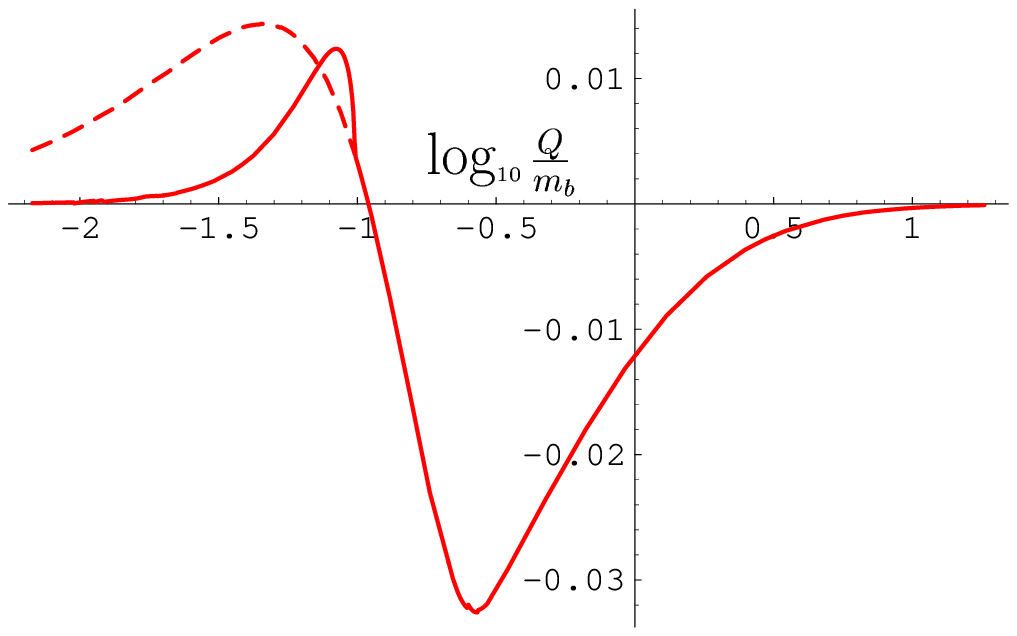,width=7.4cm}}\vspace*{-4.5mm}\\
\begin{minipage}[t]{7.4cm}
\caption{ \small
The gluon virtuality distribution for $\aver{E_x}$ without lepton
energy cut; separation scale $\mu\!=\!0.7\GeV$. 
}
\end{minipage} \hfill
\begin{minipage}[t]{7.4cm}
\caption{ \small
The same for $E_{\rm cut}\!=\!1.5\GeV$, applying  $\mu\!=\!0.45\GeV$.
}
\end{minipage} 
\end{figure}

The similar scrutiny can be applied to another short-distance component
of $\aver{M_X^2}$, the perturbative corrections to $\aver{2E_x}$. Its
naive ${\cal O}(\alpha_s)$ evaluation turns out quite small and might be
thought to be of no practical interest. However, we see from Figs.~8 and 9 
that this is a result of cancellations between softer and harder
gluons; as such it is vitiated already when running of $\alpha_s$ is
accounted for. The effect of the higher-order corrections, taken at
face value appears dramatic here. However, in the Wilsonian
approach all these corrections can be readily accounted for, and lead
only to minute changes in $\aver{M_X^2}$ for reasonably placed
cuts on $E^\ell$.

Similar in spirit and technically anatomy can, in principle, be applied
to higher hadronic moments with cuts as well, where the terms
accompanying different powers of $(M_B\!-\!m_b(\mu))$ should be analyzed
separately (say, for the second mass squared moment there are three
terms with the power from $0$ to $2$). We have to bear in mind, 
however that the higher
moments more seriously depend on the Darwin operator. Therefore,
accounting for running of $\alpha_s$ would not be too meaningful
without extending the Wilsonian treatment to higher-dimension
operators, first of all to the Darwin operator. We will present 
the corresponding results in the 
forthcoming publication \cite{future}.

\thispagestyle{plain}
\begin{figure}[hhh]\vspace*{-3.4mm}
\begin{center}
\mbox{\epsfig{file=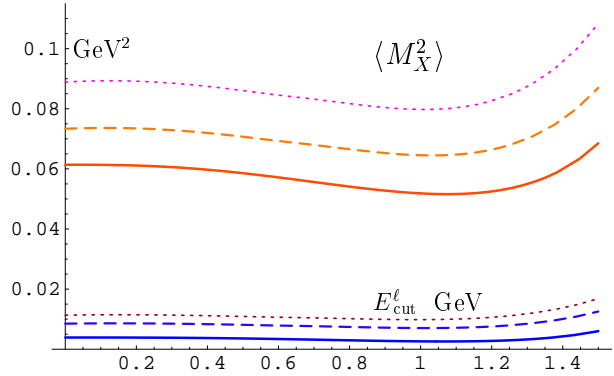,width=80mm
}}
\end{center}
\vspace*{-6.0mm}
\caption{ \small
Effect of the BLM corrections on $\aver{M_X^2}$. Upper curves show
the combined contribution: solid for the fixed-order $\alpha_s\!=\!0.3$
evaluation, dashed line for order ${\cal O}(\beta_0\alpha_s^2)$ and
dotted line at ${\cal O}(\beta_0^2\alpha_s^3)$. Lower curves are
similar effects for the contribution $\propto(M_B\!-\!m_b)$ alone.
}
\end{figure}

Here we illustrate the numerical results for $\aver{M_X^2}$. Fig.~10
shows the breakdown of the predictions for the perturbative
corrections to $\aver{M_X^2}$ as a function of the cut on lepton
energy, using the fixed-order perturbative estimates with
$\alpha_s\!=\!0.3$ \cite{slcm}, a generally appropriate choice for
beauty decays. For comparison we show also the predictions obtained by
combining the first-order ${\cal O}(\alpha_s)$ term evaluated with
admittedly too low a value $\alpha_s^{\overline{\rm MS}}\!=\!0.22$,
with the second-order BLM correction,  ${\cal O}(\beta_0\alpha^2_s)$:
\bea
\nonumber
\aver{M_X^2}= \msp{-4}&&\msp{-8} m_c^2(\mu) + (M_B\!-\!m_b(\mu))^2 \\
\nonumber
\msp{-4}&+&\msp{-4} 
(M_B\!-\!m_b(\mu))
m_b\left\{E_1^{(0)}+ C_F\left[ \mbox{$\frac{\alpha_s}{\pi} E_1^{(1)} + 
\frac{\beta_0}{2}\!
\left(\frac{\alpha_s}{\pi}\right)^2\!  E_1^{(2)} + 
\left(\frac{\beta_0}{2}\right)^2\!
\left(\frac{\alpha_s}{\pi}\right)^3\!  E_1^{(3)}+ ...$}
\right]\right\}\\
&+&\msp{-4} m_b^2\, C_F\left[\mbox{$ 
\frac{\alpha_s}{\pi} \Delta_1^{(1)} + \frac{\beta_0}{2}\,
\left(\frac{\alpha_s}{\pi}\right)^2  \Delta_1^{(2)} + 
\left(\frac{\beta_0}{2}\right)^2\,
\left(\frac{\alpha_s}{\pi}\right)^3  \Delta_1^{(3)}+ ...$}
\right]\;,\\
\nonumber
&&\msp{30} 
\alpha_s\!\equiv\!\alpha_s^{\overline{\mbox{{\tiny{\rm MS}}}\!}\,}(m_b)\,,
\qquad   \mbox{$C_F=\frac{4}{3}, \qquad 
\beta_0=\frac{11}{3}N_c-\frac{2}{3}n_f=9$} \;.
\label{50}
\eea
Such an approximation is routinely applied in lieu of a more
thoughtful choice of $\alpha_s$ where the corresponding BLM
correction is known. We can even add the higher-order BLM terms
(dotted lines in Fig.~10).

It is clear that the higher-order corrections are under good control
here. Taking the results literally, the effective value of $\alpha_s$
to be used in the fixed-order calculations is only slightly larger
than $0.3$ slowly increasing at higher cuts, and does not exceed
$0.38$, well within the interval allowed for in Ref.~\cite{slcm}.

To have an unbiased comparison, however one should recall that the part of the
shift when using the BLM-improved form is actually offset by the
change in the commensurate value of the effective Darwin
expectation value:
\beq
\tilde\rho_D^3 \:\simeq \:\left\{ \begin{array}{lll} 
\rho_D^3(1\GeV)-0.085\GeV^3 & \mbox{~fixed order, }    & \alpha_s\!=\!0.3 \\ 
\rho_D^3(1\GeV)-0.12\GeV^3  & \;\;\,{\cal O}(\beta_0\alpha_s^2), &
\alpha_s\!=\!0.22 \rule[-8pt]{0mm}{8mm}\\
\rho_D^3(1\GeV)-0.16\GeV^3  & \;\;\,{\cal O}(\beta_0^2\alpha_s^3), &
\alpha_s\!=\!0.22 
\end{array} \right.
\label{60}
\eeq
For the first-order BLM improvement this shift lies below 
theoretical accuracy and we usually neglect it, but it is relevant
when assessing the net effect of the BLM improvement. For higher orders
it becomes significant. 
\vspace*{3mm}

\noindent
\hspace*{.2em}Table 1.~~{\small Perturbative coefficients for
$\aver{M_X^2}$ at $m_c\!=\!1.16\GeV$, $m_b\!=\!4.6\GeV$, $\mu\!=\!1\GeV$}
\vspace*{1mm}\\
\begin{tabular}{|c|l|l|l|l|l|l|l|l|l|}\hline
$\! E_{\rm cut}, \GeV \! $& ~\hfill $ \Delta_1^{(1)} $\hfill~ & \hfill~
$\Delta_1^{(2)}$ \hfill~ & ~\hfill
$ \Delta_1^{(3)} $ \hfill~
& ~\hfill $\Delta_1^{(4)}$ \hfill~&~\hfill $E_1^{(0)}$ \hfill~&~\hfill $E_1^{(1)}$ 
\hfill~&\hfill~ $E_1^{(2)} $ \hfill~~&~\hfill $E_1^{(3)} $ \hfill~ &
~\hfill $\!\!E_1^{(4)}\!\!$ \hfill~ \\ \hline
$0\;\;\, $&$ 0.021 $&$ 0.036 $&$ 0.065 $&$ 0.3 $&$ 
0.846 $&$ 0.0098 $&$ 0.061 $&$ 0.097 $&$ 0.49 $  \\  \hline  
$0.6 $&$ 0.02 $&$ 0.035 $&$ 0.064 $&$ 0.3 $&$ 
0.839 $&$ 0.0085 $&$ 0.058 $&$ 0.095 $&$ 0.49 $   \\  \hline  
$0.9 $&$ 0.019 $&$ 0.034 $&$ 0.063 $&$ 0.3 $&$ 
0.827 $&$ 0.0069 $&$ 0.056 $&$ 0.095 $&$ 0.51 $   \\  \hline  
$1.2 $&$ 0.018 $&$ 0.036 $&$ 0.067 $&$ 0.33 $&$ 
0.811 $&$ 0.0069 $&$ 0.06 $&$ 0.11 $&$ 0.58 $  \\  \hline  
1.5 $ $&$ 0.023 $&$ 0.046 $&$ 0.087 $&$ 0.44 $&$ 
0.796 $&$ 0.015 $&$ 0.088 $&$ 0.15 $&$ 0.8 $  \\  \hline  
\end{tabular}\\
\vspace*{7mm}

Relegating the detailed analysis of other inclusive
semileptonic averages including higher hadronic moments, to the
dedicated publication \cite{future}, here we give a few
perturbative coefficients, including BLM corrections, for the first,
Table~1, and second, Tables~2a and 2b, hadronic mass square moments,
at different lepton energy cut. 
The coefficients for the second moment are defined in analogy to the
case of the first hadronic moment, Eq.~(\ref{50}), however they
proliferate:
\bea
\nonumber
\aver{[M_X^2-\aver{M_X^2}]^2}= \msp{-4}&&\msp{-4} (M_B\!-\!m_b)^2
m_b^2 \left\{E_2^{(0)}+ C_F\left[ \mbox{$\frac{\alpha_s}{\pi} E_2^{(1)} + 
\frac{\beta_0}{2}\! \left(\frac{\alpha_s}{\pi}\right)^2\!  E_2^{(2)} 
+ ...$}
\right]\right\}
\\
\nonumber
&\msp{-8}+&\msp{-4} (M_B\!-\!m_b)\,m_b\: C_F\left[\mbox{$ 
\frac{\alpha_s}{\pi} \Delta_{11}^{(1)} + \frac{\beta_0}{2}\,
\left(\frac{\alpha_s}{\pi}\right)^2  \Delta_{11}^{(2)} + 
\left(\frac{\beta_0}{2}\right)^2\,
\left(\frac{\alpha_s}{\pi}\right)^3  \Delta_{11}^{(3)}+ ...$}
\right]\\
&\msp{-8}+&\msp{-3} m_b^4 \;C_F\left[\mbox{$ 
\frac{\alpha_s}{\pi} \Delta_{2}^{(1)} + \frac{\beta_0}{2}\,
\left(\frac{\alpha_s}{\pi}\right)^2  \Delta_{2}^{(2)} + 
\left(\frac{\beta_0}{2}\right)^2\,
\left(\frac{\alpha_s}{\pi}\right)^3  \Delta_{2}^{(3)}+ ...$}\right]
 \;.
\label{56}
\eea
It should be stressed that higher-order coefficients are quoted only
for illustration, and should not really be added to the corrections
without carefully adjusting for the effect of the residual infrared piece.

The estimated perturbative corrections to the second
$\aver{(M_X^2\!-\!\aver{M_X^2})^2}$ and to the third
$\aver{(M_X^2\!-\!\aver{M_X^2})^3}$  hadronic mass moments depending on
the lepton cut, and their breakdown into separate terms are shown in
Figs.~2 and 3,  both in fixed-order perturbation theory.  We also 
show the results in the pole scheme ($\mu\!=\!0$) assuming, however,
the same values of the quark masses.
For numerical evaluations to order $\alpha_s^1$ we commonly use
$\alpha_s\!=\!0.3$, and assume the values of the short-distance heavy quark
masses 
$$
m_c(1\GeV)=1.16\GeV\,, \qquad m_b(1\GeV)=4.60\GeV
$$
preferred by experiment. 
\vspace*{3mm}

\noindent
\hspace*{.2em}Table 2a.~~{\small Perturbative coefficients for
$\aver{[M_X^2\!-\!\aver{M_X^2}]^2}$, in the same setting}
\vspace*{1mm}\\
\begin{tabular}{|c|l|l|l|l|l|l|l|l|}\hline
$E_{\rm cut}$& ~\hfill $ \Delta_2^{(1)} $\hfill~ & \hfill~
$\Delta_2^{(2)}$ \hfill~ & ~\hfill
$ \Delta_2^{(3)} $ \hfill~
& ~\hfill $\Delta_2^{(4)}$ \hfill~&~\hfill $\Delta_{11}^{(1)}$ 
\hfill~&\hfill~ $\Delta_{11}^{(2)} $ \hfill~~&~\hfill 
$\Delta_{11}^{(3)} $ \hfill~ & ~\hfill $\Delta_{11}^{(4)}$ 
\hfill~ \\ \hline
$0\;\;\, $&$ -0.00082 $&$ -0.0035 $&$ -0.0077 $&$ -0.046 $&$ 
-0.0023 $&$ -0.01 $&$ -0.023 $&$ -0.14 $  \\  \hline  
$0.6 $&$ -0.0016 $&$ -0.0044 $&$ -0.0091 $&$ -0.052 $&$ 
-0.0043 $&$ -0.013 $&$ -0.027 $&$ -0.16 $  \\  \hline  
$0.9 $&$ -0.0026 $&$ -0.0057 $&$ -0.011 $&$ -0.06 $&$ 
-0.007 $&$ -0.016 $&$ -0.033 $&$ -0.18 $  \\  \hline  
$1.2 $&$ -0.0036 $&$ -0.0073 $&$ -0.014 $&$ -0.073 $&$ 
-0.0099 $&$ -0.02 $&$ -0.039 $&$ -0.21 $  \\  \hline  
$1.5 $&$ -0.0046 $&$ -0.0091 $&$ -0.017 $&$ -0.089 $&$ 
-0.012 $&$ -0.024 $&$ -0.046 $&$ -0.24 $  \\  \hline  
\end{tabular}\\
\vspace*{7mm}

\noindent
\hspace*{.2em}Table 2b.~~{\small The coefficients for the term
$\propto (M_B\!-\!m_b)^2$}
\vspace*{1mm}\\
\begin{tabular}{|c|l|l|l|l|l|l|l|l|}\hline
$E_{\rm cut}  $& ~\hfill $ E_2^{(0)}$ \hfill~&~\hfill $E_2^{(1)}$ 
\hfill~&\hfill~ $E_2^{(2)} $ \hfill~~&~\hfill $E_2^{(3)} $ \hfill~ &
~\hfill $E_2^{(4)}$ \hfill~ \\ \hline
$0\;\;\, $&$ 0.194 $&$ -0.0038 $&$ -0.017 $&$ -0.029 $&$ -0.17 $  \\  \hline  
$0.6 $&$ 0.2 $&$ -0.0054 $&$ -0.02 $&$ -0.032 $&$ -0.17 $  \\  \hline  
$0.9 $&$ 0.208 $&$ -0.0076 $&$ -0.024 $&$ -0.035 $&$ -0.18 $  \\  \hline  
$1.2 $&$ 0.218 $&$ -0.01 $&$ -0.03 $&$ -0.041 $&$ -0.2 $  \\  \hline  
$1.5 $&$ 0.226 $&$ -0.013 $&$ -0.035 $&$ -0.047 $&$ -0.23 $  \\  \hline  
\end{tabular}\\
\vspace*{7mm}

The first-order perturbative corrections to the
second mass squared moment happen to be rather suppressed  near zero cut on
$E^\ell$. To some extent this is accidental and takes place just for our
choice of $\mu$ around $1\GeV$: it would be nearly complete for a
somewhat smaller $\mu$, and softer for a larger separation scale. The
BLM corrections at small $E_{\rm cut}$ then seem to have a large 
{\sf relative}
impact, nearly doubling the perturbative correction. However, as seen
from Fig.~2, the cancellation fades out at a relatively low cut,
and the significance of the BLM corrections becomes moderate already
at $E_{\rm cut}\!\gsim \!600\MeV$. Above $0.9\GeV$ the face value of
the effective $\alpha_s$ for the one-loop contribution is about $0.35$
-- even discarding the possible effect of redefining the appropriate
Darwin expectation value $\tilde\rho_D^3$. Therefore, it seems
likely that the actual perturbative corrections to the second moment
is even flatter than follows from Fig.~2.

It has been pointed out in Ref.~\cite{slcm} that, in view of possible
cancellations in the first-order perturbative coefficients, it is
unsafe to estimate the uncertainty associated with uncalculated
perturbative corrections simply as a fixed fraction of the one-loop
result. This may particularly apply to the schemes other than 
the pole one. Ref.~\cite{slcm} suggested that an additional
uncertainty should be added obtained examining certain `perturbative' 
variations of the nonperturbative expectation values, to evade such
special cases. We see that this
recipe works well here: while even a $50\%$ variation in the
perturbative corrections to the second mass moment near zero cut might
easily underestimate their actual uncertainty, considering the effect of
the Darwin operator gives the right estimate of the size, in fact 
on the safer side.

\section{Discussions}

We have calculated the one-loop perturbative corrections to all the
semileptonic decay structure functions of the heavy quarks, in the
form allowing to obtain all BLM corrections on the same footing. This
parallels the work done in Ref.~\cite{bbbsl} almost a decade ago which,
however was applicable only to the total semileptonic width. 
With the complete expressions for the power corrections to the
structure functions through order $1/m_Q^3$
available \cite{koy,grekap}, there remain no practical 
obstacles in calculating all
sufficiently inclusive semileptonic distributions with significant
precision.  As has been emphasized
in Ref.~\cite{imprec} in respect to precision determination of
$|V_{cb}|$,  now the real limitation in many cases becomes 
perturbative corrections to the Wilson coefficients of the  
power-suppressed nonperturbative operators; sometimes the effects of
the nonperturbative four-fermion averages with charm fields (called `intrinsic
charm' in Ref.~\cite{imprec}) may be noticeable.

We think that the program of precision extraction of higher nonperturbative heavy
quark parameters of $B$ mesons will be on agenda regardless of
calculation the perturbative corrections with all kinematic
constraints -- in many instances already power corrections proper lead
to significant uncertainties. In this respect studying at
$B$-factories the modified
higher hadronic mass-energy moments ${\cal N}_X^k$ detailed in
Ref.~\cite{slcm} looks promising. From the perturbative point of view 
they are similar, or even simpler than the standard moments of
$M_X^2$, and our analysis applies to them in full.

We plan to present both perturbative and
nonperturbative corrections in an (possibly cumbersome) analytic form in
the forthcoming publication \cite{future}, and to simultaneously 
provide ready-to-use computer programs making numerical evaluations an
automated procedure. 

Calculation of the perturbative $b$-decay structure functions to order
$\alpha_s^1$ is rather straightforward if only lengthy. The main
problem is to control possible typos, in particular in the computer
program which would evaluate the corrections. A single omission can both
radically change the emerging numerical result, or to be nearly
invisible numerically in a particular kinematic setting. To safeguard
our results against such complications, we have applied a number of
checks.

Firstly, a regular limit for a massless gluon was verified for all the
moments, and the logarithmic dependence of separately the virtual and
the bremsstrahlung contributions at given $q^2$ was checked to match the
known classical bremsstrahlung radiation divergence. Furthermore, the
limit $\lambda^2\!\to\! 0$ in the sum of the contributions for a
particular moment, is slow containing the terms $\propto \!\sqrt{\lambda^2}$,
rather than only those scaling like $\lambda^2$. These terms, however
are controlled by the OPE -- they reside solely in the corresponding
dependence of the pole quark masses on $\lambda^2$. Moreover, the OPE
ensures that in the first-loop corrections terms $\propto \!\lambda^2
\ln{\lambda^2}$ do not appear. This provided the serious test for the
moments -- once passing to the Wilsonian scheme at a given gluon mass
(and such a translation involves coefficients with a nontrivial
kinematic dependence), we observed a straight linear in $\lambda^2$
behavior at the small gluon mass.

At arbitrary gluon mass of order $m_c$ or below we observed
that the total width we numerically calculate precisely coincides 
with the same width we used in
Refs.~\cite{blmvcb,imprec}; those expressions were derived in
Ref.~\cite{bbbsl} using, generally, the different set of integration
variables. The total width, however depends only on $w_1$ and $w_2$,
see Eq.~(\ref{20}). We also compared the numerical results of the perturbative
corrections to the width with an arbitrary lepton energy cut
at $\lambda^2 \!\to\!0$ with the computations used previously \cite{slcm}
based on the well-established analytic expressions of Ref.~\cite{czarj}; this
verified $w_3$ as well. Together with a number of more technical cross
checks, a confidence has been gained that possible typos in the program for
evaluating moments were all eliminated. 
\vspace*{2mm}

Although perturbative effects are 
conceptually simple, getting accurate predictions requires 
thoughtful combining perturbative and nonperturbative corrections in
the optimal way. Our experience commencing with an early paper \cite{upset} 
suggests that using the literal
Wilsonian prescription which separates various contributions based on
their intrinsic momentum scale, is the way to get reliable
predictions. In return, applying it often yields accurate results 
even with minimal computational efforts, or using simplifying
approximations. 

In the present paper we have reported some results for the dependence
of the perturbative corrections to the hadronic mass moments, on the charge
lepton energy cut; an extensive analysis is in preparation
\cite{future}. 
In accord with the  expectations which
considered peculiarity of the adopted OPE implementation, we found
weak variations, for the safe intervals in $E_{\rm cut}$, well within
our estimates of the overall accuracy theory in its present
form can realistically provide \cite{slcm}. The perturbative corrections to
$\aver{M_X^2}$ actually turned out nearly constant, varying even much less
than could be anticipated. The $E_{\rm cut}$-variation in the
second mass moment, $\aver{(M_X^2\!-\!\aver{M_X^2})^2}$ is more
noticeable, consistent in sign and magnitude with the 
cut-dependent contribution from the Darwin operator. We actually found
indications that they may be further flatten when the BLM corrections are
incorporated: the latter eliminate a somewhat accidental cancellation
at very low $E_{\rm cut}$. 

It should be emphasized that such a moderate
sensitivity resulted from eliminating soft gluon effects from
the perturbative diagrams; in the schemes without such a separation (we
generically refer to them as {\tt pole}-type) the variation, as a rule,
is far more pronounced. We also find that sensible BLM improvement of
the one-loop estimates is possible within the Wilsonian approach,
although it does not seem to have dramatic impact on the a priori
safe moments. Nevertheless, we point out that there are strong
cancellations in the perturbative corrections to the hadron energy
$\aver{E_x}$ between different domains of integration. They are
vitiated once running of $\alpha_s$ is accounted for, which strongly
enhances their effect. Their significance have often been
underestimated. 

Our predictions \cite{slcm,misuse} both for the absolute values of the hadronic
moments, and for their cut-dependence were in a qualitatively good
agreement with the preliminary data reported by BaBar and CLEO. Having
calculated the full perturbative corrections, we did not find effects
which would be unexpectedly large or would show a surprising behavior;
therefore, we expect the agreement to persist, or possibly even
strengthen.  More precise verifications of the theory including fits
to many measured observables, is to follow from the dedicated
experimental analyses.

\vspace*{4mm}

\noindent
{\bf Acknowledgments:} This study would be impossible without close
collaboration with P.~Gambino, including joint work directly on the subjects
addressed here. I am especially grateful to him for his share in
obtaining a number of numerical theory predictions long awaited by
experiment, presented above. I am indebted to many experimental
colleagues from BaBar, in particular to Oliver Buchmueller, Vera Luth
and Urs Langenegger for important discussions and communications which
initiated this study, and for encouraging exchanges. It is my
pleasure to thank Ikaros Bigi for collaboration on closely related
issues and for important suggestions. I would like to thank R.~Zwicky
and G.~Uraltsev
for their help with Mathematica.
This work was supported in part by the NSF under grant number
PHY-0087419.

\end{document}